\documentclass[aps,prl,twocolumn]{revtex4-1}
\usepackage{mathptmx}

\usepackage[T1]{fontenc}
\usepackage[latin9]{inputenc}
\usepackage{float}
\usepackage{amsmath}
\usepackage{epsfig}
\usepackage{epstopdf}
\usepackage{amssymb}
\usepackage{graphicx}
\usepackage{esint}
\usepackage[normalem]{ulem}
\makeatletter
\@ifundefined{textcolor}{}
{%
\definecolor{BLACK}{gray}{0}
\definecolor{WHITE}{gray}{1}
\definecolor{RED}{rgb}{1,0,0}
\definecolor{GREEN}{rgb}{0,1,0}
\definecolor{BLUE}{rgb}{0,0,1}
\definecolor{CYAN}{cmyk}{1,0,0,0}
\definecolor{MAGENTA}{cmyk}{0,1,0,0}
\definecolor{YELLOW}{cmyk}{0,0,1,0}
}
\pdfoutput=1 
\usepackage{multirow}\usepackage{dcolumn}\usepackage{bm}\usepackage{ulem}\usepackage{amsfonts}\usepackage{color}\usepackage{colordvi}\usepackage{dcolumn}\usepackage{subfigure}
\newcommand{\beginsupplement}{%
        \setcounter{table}{0}
        \renewcommand{\thetable}{S\arabic{table}}%
        \setcounter{figure}{0}
        \renewcommand{\thefigure}{S\arabic{figure}}%
     }
\usepackage[colorlinks,bookmarks=false,citecolor=blue,linkcolor=blue,urlcolor=blue,hyperfootnotes=true]{hyperref}
\makeatother
\begin{document}
\title{\textcolor{black}{Angle Resolved Photo-Emission Spectroscopy signature of the Resonant Excitonic State}}
\date{\today}
\author{ X.\ Montiel, T.\ Kloss, C.\ P\'{e}pin}

\affiliation{IPhT, L'Orme des Merisiers, CEA-Saclay, 91191 Gif-sur-Yvette, France }
\begin{abstract}
We calculate the Angle Resolved PhotoEmission Spectroscopy (ARPES) signature of the Resonant Excitonic State (RES) that was proposed as the Pseudo-Gap state of cuprate superconductors [ArXiv 1510.03038]. This new state can be described as  a set of excitonic (particle-hole) patches with an internal checkerboard modulation. Here, we modelize the RES as a charge order with  $\bf{2p_{F}}$ wave vectors, where $\bf{2p_{F}}$ is the ordering vector connecting two opposite sides of the Fermi surface. We calculate the spectral weight and the density of states in the RES and we find that our model correctly reproduces the opening of the PG in Bi-2201.
\end{abstract}
\maketitle
\textit{Introduction}
The recent discovery of charge density wave (CDW) in cuprate compounds developing at the tips of the Fermi arcs in the Pseudo-Gap (PG) phase \cite{Alloul89,Warren89,Tallon01}, questions us on the relationship between $d$-wave superconductivity (SC) and the charge sector\cite{Hoffman02,Hanaguri04,DoironLeyraud07,Sebastian12,Doiron-Leyraud13,Wise08,Fujita12,He14,Tabis14,Blanco-Canosa14,Fujita14,Wu11,Wu14,Wu13a,Wu13,Fink09,Wu:2015bt,Tabis14,Ghiringhelli12,Achkar12,Comin:2015vc,Comin:2015ca,daSilvaNeto:2014vy,Comin14,Fujita29072014,Tranquada08,Tranquada09,He11,Xia08,Xia14,Hashimoto10,Sidis10nat,Lubashevsky14,Ando02,Hinkov07}.  Recent Scanning Tunneling Microscopy (STM) experiments have shown that the intrinsic energy scale of the $d$-wave part of the CDW is comparable to the PG scale \cite{Hamidian15a}. In this paper we discuss Angle Resolved Photo-Emission Spectroscopy (ARPES)  in Bi-2201 \cite{Yoshida:2012kh,Hashimoto14}.  In this compound, the experimental dispersion below the PG temperature $T^{*}$,  shows a very unusual structure in momentum space, with a  a bending of the dispersion at wave vectors very similar to the CDW  modulation vectors $\bf{Q}_x$ and $\bf{Q}_y$ \cite{Yoshida:2012kh,Hashimoto14}.  It is one of the very few experimental evidences \cite{Hamidian15a} for a correlation between a modulation structure and the  opening of the PG in the anti-nodal (AN) region ($(0,\pm \pi)$, $(\pm \pi, 0)$) of the first Brillouin zone (BZ). For most of the bulk probes, the  CDW as a coherent, static modulation shows up at a critical temperature lower than $T^*$.

Various theoretical approaches have been proposed to explain the physics of cuprate and the PG state based on strong correlations \cite{Lee06,Gull:2013hh,Sorella02}, antiferromagnetic (AF) fluctuations \cite{abanov03,Norman03,sfbook}, loop current \cite{Kotliar90,Varma97} or emergent symmetry models \cite{Zhang97,Demler04}. From these different approaches, several models describe the competition between $d$-wave SC and incommensurate CDW order \cite{Castellani95,Castellani96,Chakravarty01,Metlitski10a,Efetov13,Meier14,Greco11,Hayward14,Sachdev13,Wang14,Tsvelik14,Melikyan14,Chowdhury14,Chowdhury:2014cp,Pepin14,Wang15a,Atkinson15,Wang15c,Kloss15}, or pair density waves (PDW)- a SC state with finite Cooper pair momentum phase- \cite{Fradkin:2015ch,Lee14,Fradkin12,Agterberg:2014wf,Senthil09,Corboz:2014ba} or with current loops \cite{Kotliar90,Varma97,Bulut:2015jt}.
The emergent symmetry scenario explains the PG phase by a composite SC and charge order parameter with an underlying SU(2) symmetry which is restored by thermal fluctuations \cite{Metlitski10b,Efetov13}. Proceeding by the integration over the SU(2) pairing fluctuations \cite{Kloss15a}, we find a new stable state which manifests itself as a set of excitonic (particle-hole pair)  patches with an internal structure of checkerboard charge modulation which we call "Resonant Excitonic State" (RES). This new state is a superposition of charge density wave instabilities with multiple ordering vectors that are commensurate with the Fermi surface.

\begin{figure}
\includegraphics{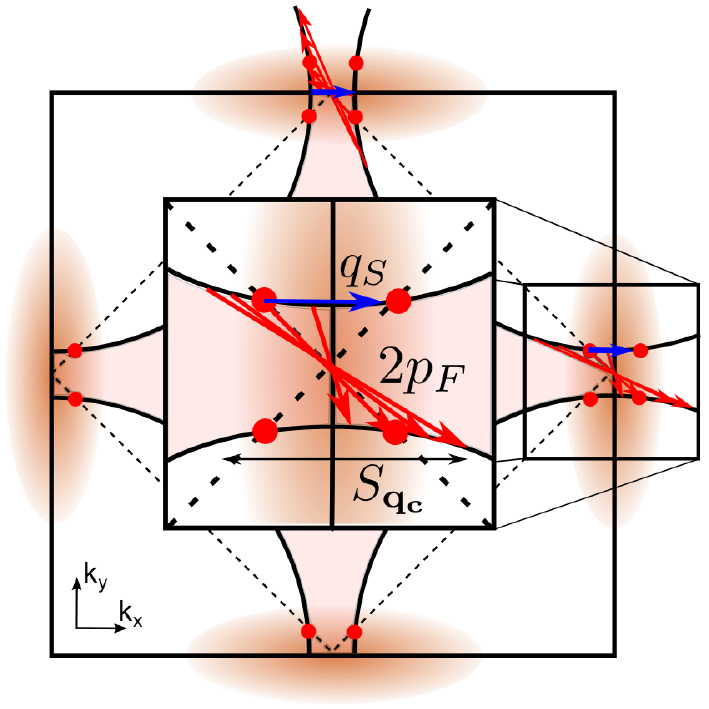}
\caption{\label{figFermi}(Color online) Fermi surface (solid line) in first BZ of the square lattice. The magnetic BZ is presented in dashed line and its intersection with the Fermi surface represents the position of the hot-spots (the red circle). The red solid arrow represents the $\bf{2p_{F}}$ vectors for two different points on the Fermi surface. The blue arrow represents the small ordering vector perpendicular to the zone edge $\bf{q_{S}}$, which is itself a $\bf{2 p_F}$- vector of the adjacent AN zone. The RES order parameter develops in the AN zone of the first BZ (orange area). $S_{\mathbf{q_{c}}}$ is the size of the spread of the modulation vectors $\mathbf{q_{c}}$.}
\end{figure}

In this paper, we propose a minimal model that describes RES as a charge order with only two kinds of ordering vectors (see Fig.\ref{figFermi}). Then, we examine carefully the gapping out of the AN region, and show that our results are in promising agreement with the experimental band structure observed by ARPES in Bi-2201\cite{Yoshida:2012kh,Hashimoto14}.

\begin{figure*}
\includegraphics{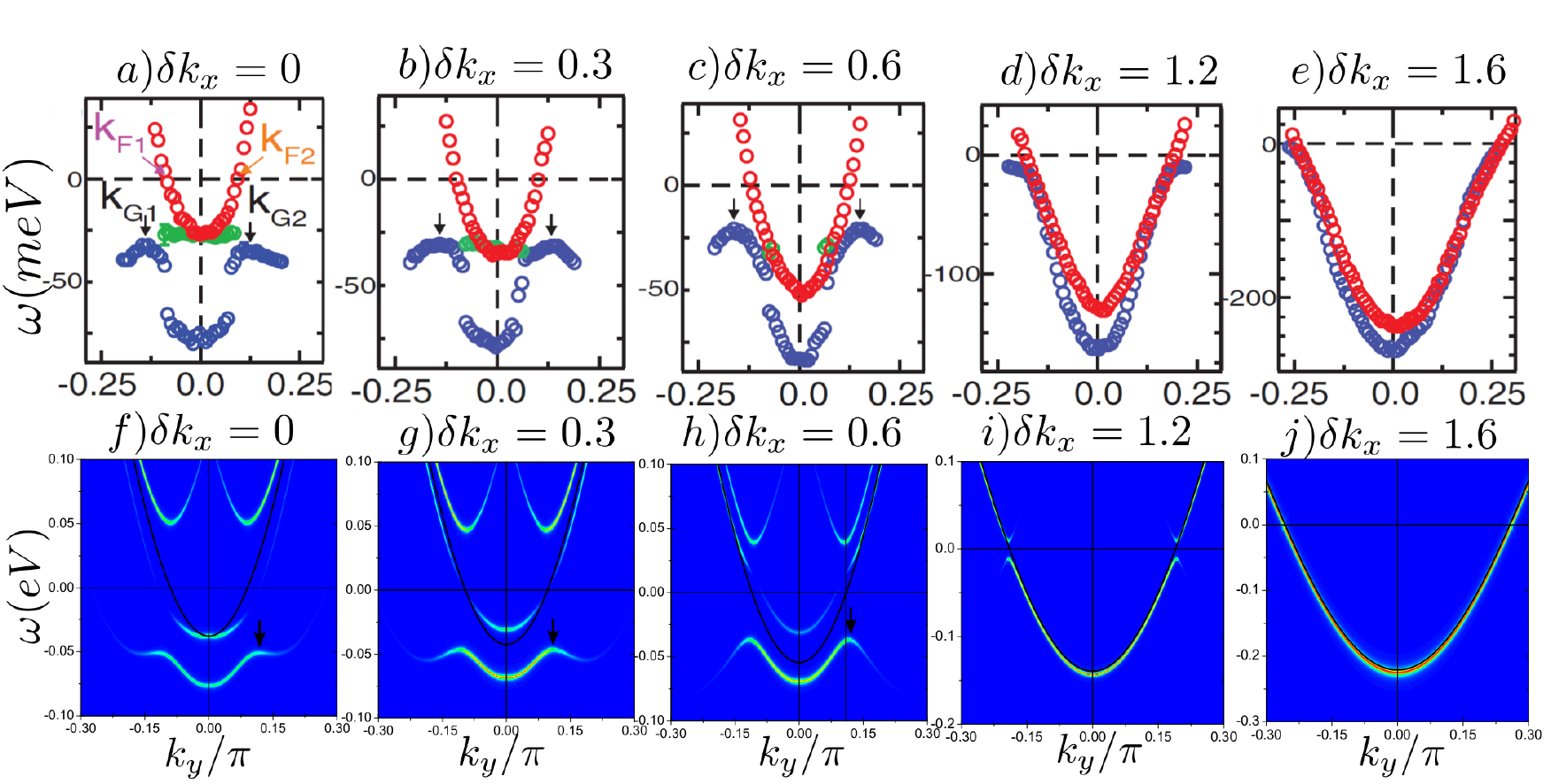}
\caption{\label{allfig}(Color online) a) to e) Experimental energy dispersion curves that were taken from \cite{He11}  and measured by ARPES at $T=10K$ (blue circle) and $T=170K$ (red circle) at approximately constant $k_{x}=\pi-\delta_{k_{x}}$ for $\delta_{k_{x}}=0,\,0.3,\,0.6,\,1.2,\,1.6$ (The Fermi arcs begin around $\delta_{k_{x}}=1.2$). Note that the green circles appear only below $T_{c}$. The black arrows locate the back-bending vector $k_{G}$ which differ from the Fermi momentum in the normal state $k_{F}$. In the figures f) to j), we represent the band dispersion with color intensity proportional to the spectral weight $A(\mathbf{k},\omega)$. These curves have been calculated for the order parameters magnitude $\Delta_{0}^{\bf{2p_{F}}}=50meV$ and $\Delta_{0}^{\mathbf{q_{S}}}=0.4\Delta_{0}^{\bf{2p_{F}}}$ and the width of the Gaussian function as $\sigma_{x}=0.5414$ and $\sigma_{y}=1.083$. The solid lines is the electronic dispersion in the metallic phase.}
\end{figure*}

ARPES technique gives crucial information about the momentum and frequency dependence of the one-particle spectral function below the Fermi level. Numerous experiments have been done in the PG state in different cuprate compounds \cite{Hashimoto14,Yoshida:2012kh,Vishik:2012cc,Vishik:2010tc,Vishik14}. In Bi2201, the electronic dispersion in the PG phase presents the following characteristics \cite{He11,Yoshida:2012kh}:

(1) The back-bending vectors of the hybridized band $k_{G}$ is greater than the Fermi vector in the normal state $k_{F}$ ($k_{F}$ and $k_{G}$ are drawn in Fig.\ref{allfig} a)). The misalignment $k_{G}-k_{F}$ develops below $T^{*}$, the PG critical temperature, in the antinodal zone and becomes smaller when moving from the AN zone until the beginning of Fermi arcs\cite{He11,Yoshida:2012kh}. 

(2) When moving from the AN zone until the Fermi arc, the hybridized band approaches the Fermi level from "below" i.e. from negative energies. 

(3) The minimum energy of the hybridized band is smaller than the minimum energy of the bare dispersion in the metallic phase. 

(4) The hybridized band is not flat and conserves a dispersion close to the dispersion of the metallic phase in the AN zone.

\textcolor{black}{The misalignment $k_{G}-k_{F}$ (feature 1) has first been interpreted by the presence of an ordered state like CDW. However, it has been emphasized \cite{Lee14,SI} that a CDW order should open a gap at only one point of the FS and that the gap should close from "above" i.e. from high energies when moving from the AN zone until the N zone which contradicts the experimental observations (feature 2) \cite{He11,Yoshida:2012kh}. In this framework, the features (1) and (2) have been explained by a PDW state that emerges at the same wave vector than the CDW state \cite{Wang14,Lee14}. Here, we argue that the RES-scenario  gives a viable alternative to the previous explanation of ARPES data in terms of a PDW.}

\textit{Model}
The RES is described as a superposition of charge orders with a multiple set of ordering vectors $\mathbf{q_{c}}$, each having the amplitude $\Delta_{RES}$ and a finite width of action in $\bf{k}$-space. Consequently, the RES order parameter depends on both $\mathbf{q_{c}}$ and on the momentum $\mathbf{k}$. $\Delta_{RES}^{\mathbf{q_{c}}}(\mathbf{k})\equiv \langle c^{\dagger}_{\mathbf{k+q_{c}},\sigma}c_{\mathbf{k},\sigma}\rangle$ with $c^{(\dagger)}_{\mathbf{k},\sigma}$ the annihilation (creation) operator of a spin $\sigma$ electron with momentum $\mathbf{k}$.

\textcolor{black}{In real space, the RES order parameters writes :}
\begin{align}
\Delta_{RES}(\mathbf{r},\mathbf{r'})=\sum_{\mathbf{q_{c},k}}e^{-i\mathbf{q_{c}}.\frac{\mathbf{r}+\mathbf{r'}}{2}}e^{i\left(2\mathbf{k}+\mathbf{q_{c}}\right)\left(\mathbf{r}-\mathbf{r'}\right)} \Delta_{RES}^{\mathbf{q_{c}}}(\mathbf{k}) 
\end{align}
where the ordering vectors $\mathbf{q_{c}}$ runs over the $\mathbf{2p_{F}}$ vectors \cite{Kloss15a}. The order parameter $\Delta_{RES}(\mathbf{r},\mathbf{r'})$ describes a local patch, made of particle-hole pairs that break locally the translation invariance.
In analogy with Cooper pairs, these particle-hole pairs have a coherence length $\xi \approx \hbar v_{F}/|\Delta_{RES}|$. The coherence lenght is about $a_{0}$  where $a_{0}$ is the elementary cell parameter of the square lattice (with $|\Delta_{RES}|$ the typical amplitude of the RES gap in the AN zone of the first BZ and $v_{F}$ the Fermi velocity). The typical size of a RES patch $L_{RES}$ depends on the spread of the $\mathbf{q_{c}}$ vectors, $S_{\mathbf{q_{c}}}$. Then, $L_{RES}\approx 2\pi\hbar/S_{\mathbf{q_{c}}}$ and is around $5a_{0}$ in agreement with STM experiments \cite{Hamidian15a}.
Finally, each patch exhibits an internal charge modulation resulting from the sum over the ordering vectors $\mathbf{q_{c}}$. These modulation has a checkerboard form along the x and y axis (more informations in Refs. \cite{Kloss15a}).  Another scenario involving CDW with multiple ordering vectors has recently been proposed to explain the ARPES data \cite{Harrison14,Kivelson2016}.

The exact resolution of the RES mean field equation \cite{Kloss15a} reveals that two sets of ordering vectors mainly contribute: $\mathbf{2p_{F}(k)}$ and $\mathbf{q_{S}}$ (see Fig. \ref{figFermi}). In the following, we neglect the contribution of the other ordering vectors. The $\mathbf{2p_{F}(k)}$ vectors relate two opposite part of the Fermi surface (red vectors in Fig. \ref{figFermi}). We assume that the $\bf{2p_{F}}$ vector of a point far from the Fermi surface is the $\bf{2p_{F}}$ vector of the closest point of the Fermi surface. The $\mathbf{q_{S}}$ vectors relate to points of the same Fermi surface across the zone edge (blue vectors in Fig. \ref{figFermi}). The $\mathbf{q_{S}}$ vector corresponds to the  $\bf{2p_{F}}$ vector of the zone edge point of the adjacent AN zone. The ordering vector $\mathbf{q_{S}}$ is a subsidiary contribution compared to the $\bf{2p_{F}}$ ordering vector. This implies that the magnitude of the order parameter for $\mathbf{q_{S}}$ ordering vectors is smaller than the $\bf{2p_{F}}$ one, $\Delta_{RES}^{2p_{F}}>\Delta_{RES}^{q_{S}}$.

The second approximation is that we neglect the detailed momentum dependence of the RES order parameter associated with each ordering vectors. Note that the exact resolution of the mean field equation gives similar momentum dependence for the $2\mathbf{p_{F}(k)}$ and $\mathbf{q_{S}}$ RES order parameters \cite{Kloss15a}. 

Following these approximations, we can describe the RES by the effective action $S_{eff}$ which writes in the basis $\Psi^{\dagger}_{\bf{k}}=\left(c^{\dagger}_{\mathbf{k},\sigma},c^{\dagger}_{\mathbf{k+2p_{F}(k)},\sigma},c^{\dagger}_{\mathbf{k+q_{S}},\sigma}\right)$ with $S_{eff}=-\sum_{\bf{k},\sigma}\psi_{\bf{k}}^{\dagger}\hat{G}_{eff}^{-1}\Psi_{\bf{k}}$ and:
\begin{equation}
\hat{G}_{eff}^{-1}=\left(\begin{array}{ccc}
i\epsilon-\xi_{\mathbf{k}} &\Delta^{2\mathbf{p_{F}}}_{RES}(\mathbf{k})&\Delta^{\mathbf{q_{S}}}_{RES}(\mathbf{k})\\
\Delta^{2\mathbf{p_{F}},*}_{RES}(\mathbf{k})& i\epsilon-\xi_{\mathbf{k+2p_{F}(k)}}&0\\
\Delta^{\mathbf{q_{S}},*}_{RES}(\mathbf{k})&0& i\epsilon-\xi_{\mathbf{k+q_{S}}}\\
\end{array}\right)\label{Mat3x3}\ ,
\end{equation}
where $\epsilon$  is the fermionic Matsubara frequency and $\xi_{\bf{k}}$ the fermionic spectrum \footnote{We use a tight-binding approximation to describe the electronic spectrum with
$\xi_{{\bf {k}}}=t_{1}\left[\text{cos}(k_{x}a_{0})+\text{cos}(k_{y}a_{0})\right]
+2t_{2}\text{cos}(k_{x}a_{0}\text{cos}(k_{y}a_{0})
+t_{3}\left[\text{cos}(2k_{x}a_{0})+\text{cos}(2k_{y}a_{0})\right]
+t_{4}\left[\text{cos}(2k_{x}a_{0})\text{cos}(k_{y}a_{0})+\text{cos}(k_{x}a_{0})\text{cos}(2k_{y}a_{0})\right]-\mu$
where $t_{i}$ are the hopping parameter to the first ($i=1$) until the $i^{th}$ nearest neighbour (where $i=1..4$) and $\mu$ is the chemical potential. In the following, we put $a_{0}$ to unity, $a_{0}=1$. The value of the hopping parameters are $t_{1}=-0.44\,eV$, $t_{2}=0.06863\,eV$, $t_{3}=-0.071954\,eV$ and $t_{4}=-0.0286548\,eV$ and $\mu=-0.24327\,eV$.}.  We choose a fermionic spectrum that reproduce the dispersion of the Bi-2201 compounds at optimal doping \cite{He11}.

We assume that the RES order parameter has a Gaussian form centered in the AN zone of the first BZ (see orange area in Fig. \ref{figFermi}) $\Delta^{\mathbf{q_{c}}}_{RES}(\mathbf{k})=\Delta^{\mathbf{q_{c}}}_{0}\text{e}^{-\frac{(k_{x}-\pi)^{2}}{2\sigma_{x}^{2}}-\frac{(k_{y})^{2}}{2\sigma_{y}^{2}}}$ where $\sigma_{x(y)}$ is the width of the Gaussian in the $k_{x}(k_{y})$ direction.
In order to reproduce the experimental dispersion measured by ARPES, we calculate the electronic spectral weight $A(\mathbf{k},\omega)$ which defines the probability to find the state $c_{k,\uparrow}$ in the eigenstate of the effective Hamiltonian. The spectral weight can be written in term of Green function as
$A(\mathbf{k},\omega)=-2 \text{lim}_{\eta \rightarrow 0}\text{Im}\left(G^{1,1}(\mathbf{k},\omega+i\eta)\right)$
where $i\epsilon=\omega+i\eta$ and the Green function $G^{1,1}(\mathbf{k},i\epsilon)$ is obtained by the inversion of the matrix (\ref{Mat3x3}) and reads 
\begin{align}
G^{1,1}(\mathbf{k},i\epsilon)=-\frac{(i\epsilon-\xi_{\mathbf{k+2p_{F}(k)}})(i\epsilon-\xi_{\mathbf{k+q_{S}}})}{\text{det}(\hat{G}_{eff}^{-1})}.
\end{align}
where $\text{det}(\hat{G}_{eff}^{-1})$ is the determinant of the matrix $\hat{G}^{-1}_{eff}$. 
In the following, we use a broadening $\eta=3\,meV$ which is small compared to the bandwidth and the amplitude of the order parameters.

\textit{Results and discussion} In the figure \ref{allfig} a) to e), we present the experimental dispersion measured with ARPES (extracted from \cite{He11}) in order to compare it with the theoretical dispersion (Fig.\ref{allfig} f) to j)). The magnitude of the order parameter are $\Delta_{0}^{\bf{2p_{F}}}=50\,meV$ and $\Delta_{0}^{\bf{q_{S}}}=0.4\Delta_{0}^{\bf{2p_{F}}}$ and we choose $\sigma_{x}=0.5414$ and $\sigma_{y}=1.083$. These parameters were obtained by a fit of the gap function obtained from the solution of the mean field equation \cite{Kloss15a}. Close to the zone edge, we distinguish two distinct bands below the Fermi level  (Fig.\ref{allfig} f) g) h)): the lower energy band (Fig.\ref{allfig} f) g) h)) well fits the experimental data (blue dots in Fig.\ref{allfig} a),b),c)); the middle energy band (Fig.\ref{allfig} f) g) h)) well reproduces the shoulder observable experimentally (green dots in Fig.\ref{allfig} a),b) c)). Since the shoulders observed in EDC's curve are observable only below $T_{c}$, we argue that it could not be observable above $T_{c}$ because of the loss of sensibility of the ARPES techniques with temperature.

Close to the Fermi arcs, we see a good agreement between experimental (Fig.\ref{allfig} d) e)) and calculated (Fig.\ref{allfig} i) j)) spectral weight. Following the experimental data, the gap is closing from below when $k_{x}$ varies from the AN zone ($\delta k_{x}=0$) until the nodal zone ($\delta k_{x}=1.6$).\textcolor{black}{The RES does not gap all states of the first BZ but only states in the AN zone. Consequently, the gap closes when we move from the AN zone until the Fermi arcs.}

The charge ordering vectors $\bf{2p_{F}}$ and $\bf{q_ {S}}$ impose that the gap opens at the Fermi surface with a misalignment of the back-bending curve with the Fermi vector in the normal state. This effect of misalignment is well reproduce by the simplified model (Fig. \ref{allfig} f) to j)) and decreases close to the Fermi arc (black arrow in Fig \ref{allfig} f) g) h).

\begin{figure}
\includegraphics{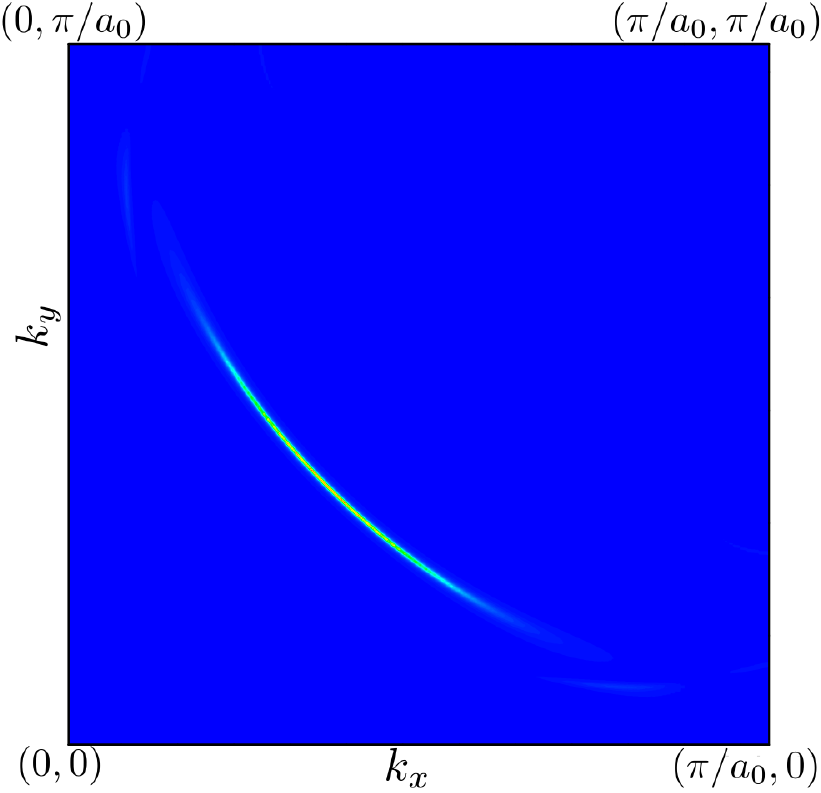}
\caption{\label{FS}(Color online) The spectral weight $A(\mathbf{k},\omega)$ at the Fermi level (at $\omega=0$ with a broadening of $\eta=10\,meV$). The parameters are the same as in Fig. \ref{allfig}. We clearly observe the formation of Fermi arc after the anti-nodal zone has been gapped out. }
\end{figure}

\begin{figure}
\includegraphics{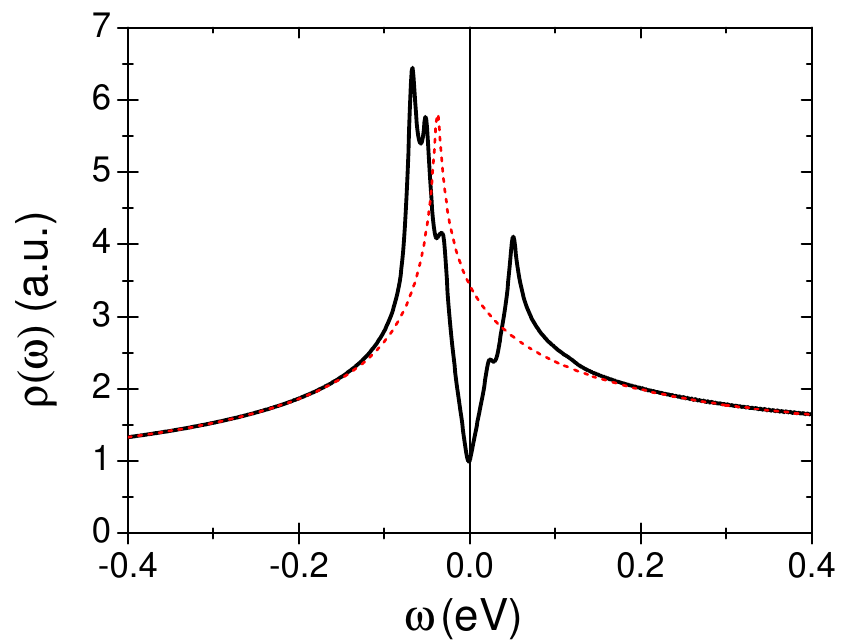}
\caption{\label{denss}(Color online) Density of states $\rho(\omega)$ as a function of the frequency in the PG state (solid line) and in the normal state (dashed line). The parameters are the same than in Fig. \ref{allfig}.}
\end{figure}

The Fermi surface of the effective model is shown in Fig. \ref{FS}. The gap develops only in the antinodal zone which implies that the Fermi surface is reduced to Fermi arcs. Note that there is no emergence of electron pockets on the axis contrary to what is found with PDW phase \cite{Lee14}.
The density of states $\rho(\omega) = \frac{1}{\pi}\sum_{\mathbf{k}}A(\mathbf{k},\omega)$  of the RES is presented in Fig. \ref{denss}. We observe the opening of a pseudo gap which qualitatively reproduce the STM data \cite{Hamidian15a}. 
Our simple model provides promising agreement with the experimental spectrum.

In experiments, the gap develops generically in the AN zone and is characteristic for the PG state. With our model, we propose a gap dependence with a Gaussian form centered in the AN of the FBZ which allows the formation of Fermi arcs which well modelizes the ARPES data \cite{He11,Yoshida:2012kh}. \textcolor{black}{Moreover, the hybridized band produced in our model conserves its dispersion in the AN zone.} In a model describing a CDW ordered state with a unique ordering vector, the hybridized bands become flat if the order parameter magnitude is comparable to the band width (see discussion in \cite{Lee14}). Consequently, the conservation of the electronic dispersion in the AN zone observed in the PG phase of cuprate compounds \cite{He11,Yoshida:2012kh} is in favour of a scenario of a charge ordered state with multiple ordering vectors.

Another feature is the misalignment between the Fermi vector in the metallic phase $k_{F}$ and the back-bending vector in the PG phase $k_{G}$. This misalignment is greater in the AN zone and decreases close to the Fermi arcs \cite{He11,Yoshida:2012kh}. It suggest the presence of an order parameter describing a state with a non-zero ordering vector such as CDW or PDW order \cite{He11} . \textcolor{black}{In our scenario, the misalignment $k_{G}-k_ {F}$ is produced by the ordering vector $\mathbf{q_{S}}$}. Note that this misalignment can also be produced in a charge order model with more than two ordering vectors.

When moving from the AN zone until the N zone, we reproduce the closure of the gap from negative energies (from "below") (Fig \ref{allfig} f) to j)). 
\textcolor{black}{Former interpretations claimed the impossibility to reproduce this gap dependence with a CDW order. The reason invoked was that a CDW phase should open a gap below the Fermi level and at only one point of the FS for a careful choice of ordering vector \cite{Lee14}. In our model, the closure of the gap from negative energies  is due to the {\it finite extension} in $\bf{k}$ of the multiple wave vector CDW state that develops only in the AN zone and close to the FS.}

\textcolor{black}{To sum up, the RES can be modelized as a CDW state with multiple ordering vector developing only in the AN zone. The closure of the gap from negative energies observed in experiments \cite{He11,Yoshida:2012kh} is well reproduced because the RES only develops in the AN zone. Moreover, the misalignment between the Fermi vector in the normal state and the back-bending vector of the hybridized band \cite{He11,Yoshida:2012kh} can be reproduced because of the second ordering vector $\bf{q_{S}}$ \cite{SI}. The RES does not produce flat hybridized bands in promising agreement with experiments \cite{He11,Yoshida:2012kh}.}

\textcolor{black}{With the temperature, we expect the gap magnitude to decrease and then to observe the closure of Fermi arcs as well as the decreasing of the misalignment $k_{G}-k_ {F}$ \cite{SI}. No qualitative evolution is expected with hole doping, because the ordering vectors of the RES change with the FS.}

The minimal model produces two hybridized bands below the Fermi level which are in promising agreement with ARPES data \cite{He11,Yoshida:2012kh}. Note that the middle energy hybridized band qualitatively reproduce the shoulder oberved in the EDC below $T_{c} $\cite{He11,Yoshida:2012kh}. In our model, we observe this band above $T_{c}$. We argue that it is a limitation of the minimal model where we modelize the RES by a three band effective model.

\textit{Conclusion}
The results presented clearly demonstrate the possibility to reproduce qualitatively the electronic spectrum of Bi-2201 observed by ARPES \cite{He11,Yoshida:2012kh}.  Our model, the RES, is a charge order scenario,  which partially contradict former interpretations Ref.\cite{Lee14} . The superposition of multiple ordering wave vectors in each excitonic, particle-hole patch, as well as the finite with of action of each wave vector within $\bf{k}$-space are the key ingredients for the successful description of the data. The RES develops in the AN zone of the first BZ \cite{Kloss15a} and is responsible for the PG opening.

\begin{acknowledgments}
The authors acknowledge Y. Sidis for helpful discussions. This work was supported by LabEx PALM (ANR-10-LABX-0039- PALM), the ANR project UNESCOS ANR-14-CE05-0007, as well as the Grant No: Ph743-12 of the COFECUB which enabled frequent visits to the IIP, Natal. X.M. and T.K. also acknowledge the support of CAPES and funding from the IIP.
\end{acknowledgments}

\newpage
\bibliographystyle{apsrev4-1}
\bibliography{Cuprates}
\clearpage
\onecolumngrid
\newpage
\part*{Supplemental Material}
\beginsupplement
\begin{section}{Qualitative evolution of the electronic dispersion with gap magnitude}
In this section, we study the evolution of the electronic band structure at the zone edge ($k_{x}=\pi$) with the order parameter magnitude. In a mean field treatment of the RES order parameter, one expects the  order parameter magnitude of the RES to decrease with the temperature.  Note that we neglect the $d$-wave superconducting (SC) order parameter in our study. Consequently, the results presented here are relevant for temperature above $T_{c}$ and below $T^{*}$ ($T_{c}$ is the SC critical temperature and $T^{*}$ the Pseudo-gap (PG) critical temperature).

\textcolor{black}{For $T_{c}<T<T^{*}$, the ARPES data show that the gap closes with the temperature and vanishes at $T^{*}$. Moreover, the misalignment $k_{G}-k_{F}$ decreases when the temperature increases. Below $T_{c}$ a shoulder has been detected in the Energy Distribution Curves (EDCs) which could be interpreted as an additional band \cite{He11}.}

With our model, we see two effects of the decreasing of the gap magnitude (see Fig. \ref{gaptemp}). First, the order parameter at the Fermi surface decreases and the Fermi arcs closes (see Fig. \ref{gaptemp} e) to h)) which is qualitatively coherent with observations \cite{He11,Yoshida:2012kh}. The second effect is that the spectral weight of the low energy band decreases in favour of the middle energy band (see Fig. \ref{gaptemp}a) to d)). The disappearance of the low energy band in favour of the middle energy band implies that the misalignment $k_{G}-k_{F}$ decreases until vanishing at $T^{*}$ where the RES is expected to disappear. Note that below $T_{c}$, the variation of the RES order parameter amplitude with temperature can be non-trivial because of the appearance of the superconductivity.
\begin{figure}[!h]
\includegraphics{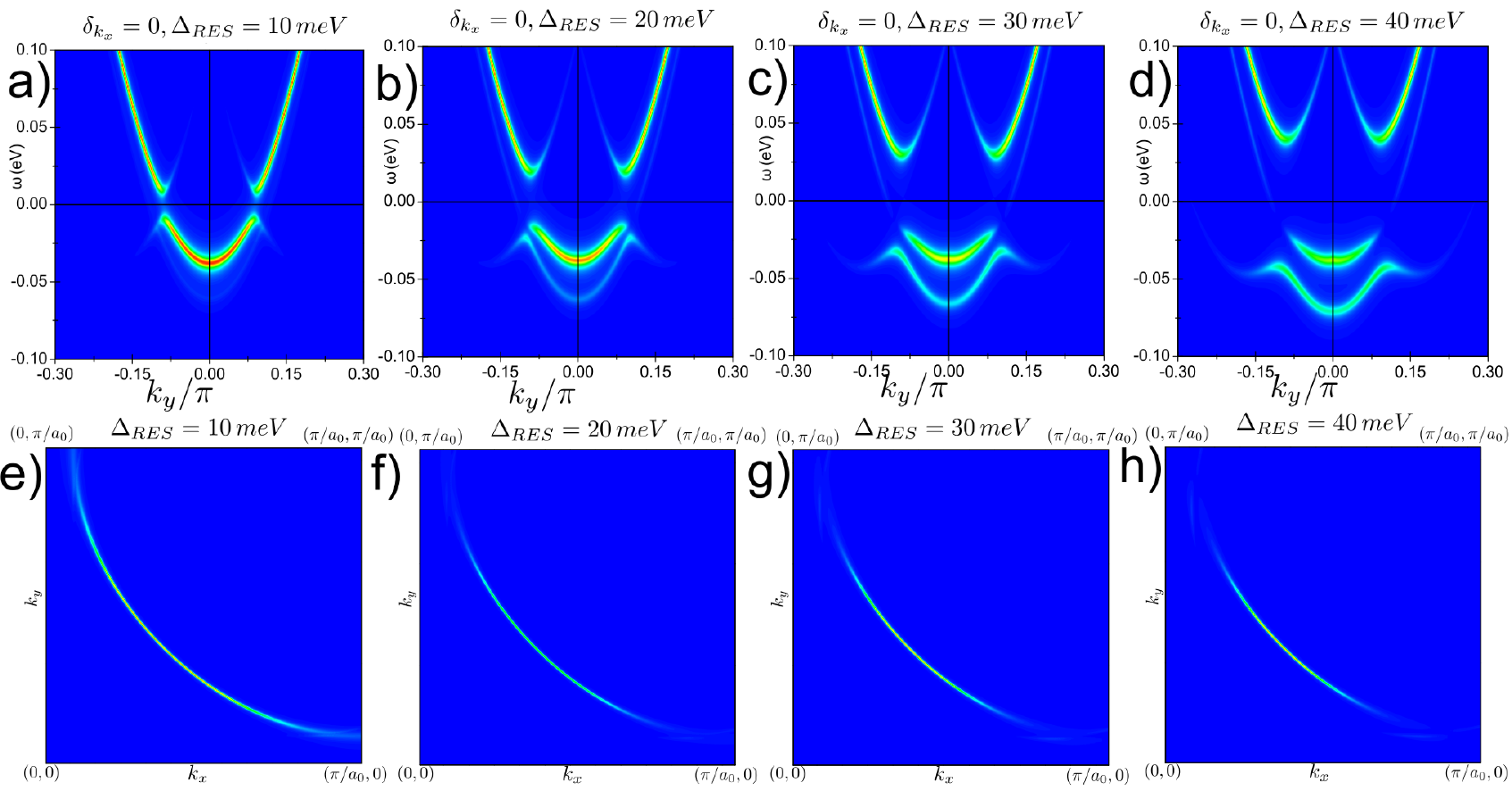}
\caption{\label{gaptemp}(Color online) Spectral weight $A(\mathbf{k},\omega)$ at the zone edge ($\delta_{k_{x}}=0$) for a gap amplitude $\Delta_{RES}$ equals to a) $10 \, meV$ b) $20\, meV$ c) $30\, meV$ and d) $40 \,meV$ (with $\Delta_{0}^{\bf{2p_{F}}}=\Delta_{RES}$ and $\Delta_{0}^{\mathbf{q_{S}}}=0.4\Delta_{RES}$). The spectral weight at the Fermi level $A(\mathbf{k},\omega=0)$ is presented for a order parameter magnitude equals to e) $10 \, meV$ f) $20\, meV$ g) $30\, meV$ and h) $40 \,meV$. We clearly see the Fermi arcs close when the order parameter magnitude decreases.}
\end{figure}
\end{section}
\newpage
\begin{section}{Solution of the minimal model for an arbitrary Fermi surface}
In this part, we  demonstrate that our approach does not depend on the Fermi surface topology. In this case, we consider a Fermionic dispersion $\xi_{\mathbf{k}}$ that writes $\xi_{\mathbf{k}}=-2t(\cos(k_{x}a_{0})+\cos(k_{y}a_{0}))+4t'\cos(k_{x}a_{0})\cos(k_{y}a_{0})+t_{0}(\cos(k_{x}a_{0})-\cos(k_{y}a_{0}))^{2}-\mu$ where $t,t'$ are respectively the first and second neighbor hopping terms with $t'=-0.3t$ and $t_{0}=0.084t$  take into account the interlayer coupling. $a_{0}$  is the elementary cell parameter set to unity and $\mu$ is the chemical potential determined to adjust the hole doping. The bandwidth parameter has been put at $1.5\,eV$. The band structure (Fig. \ref{bandqqc}) as well as the Fermi surface (Fig. \ref{gapqqc}) behave in a similar manner than the one presented in the Fig. \ref{allfig} and \ref{FS} in the main paper.
\begin{figure}[!h]
\includegraphics{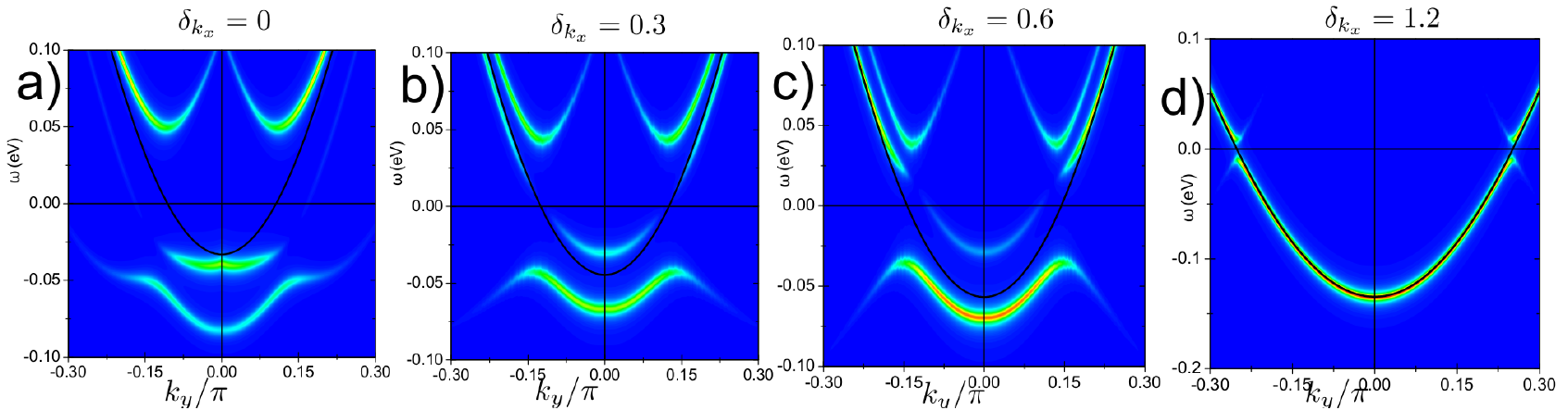}
\caption{\label{bandqqc}(Color online) Spectral weight $A(\mathbf{k},\omega)$ as a function of the frequency for a CDW order with $\bf{2p_{F}}$ and $\bf{q_{S}}$ ordering vectors at a) $\delta_{k_{x}}=0$, a) $\delta_{k_{x}}=0.3$, a) $\delta_{k_{x}}=0.6$ and a) $\delta_{k_{x}}=1.2$. We clearly see a gap opening in the AN zone. The magnitude as well as the form factor of the order parameters are the same than in Fig. \ref{allfig}.}
\end{figure}
\begin{figure}[!h]
\includegraphics{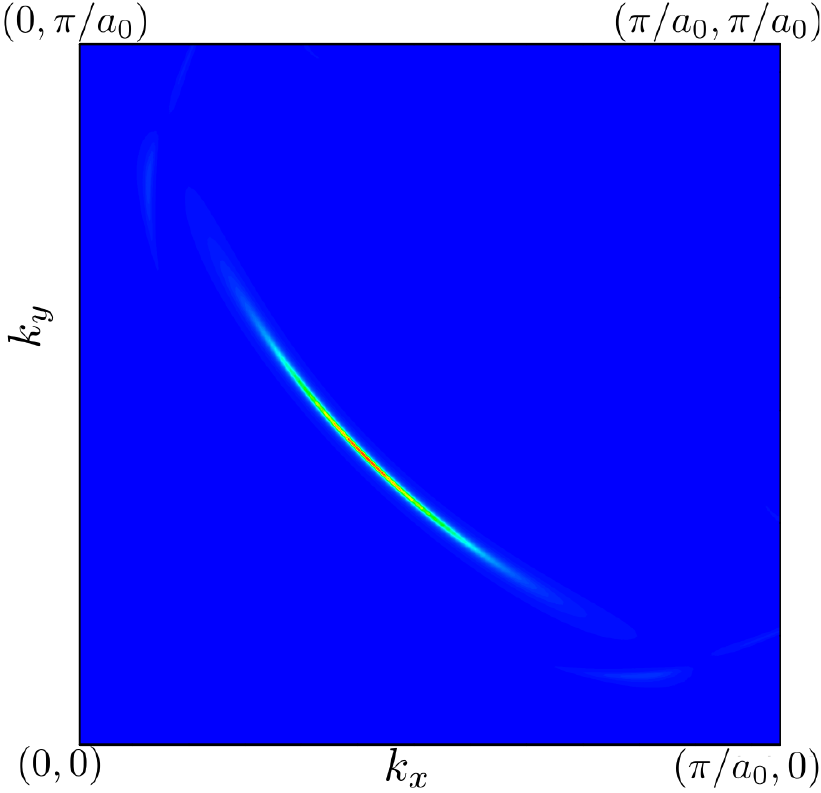}
\caption{\label{gapqqc}(Color online) Spectral weight at the Fermi level $A(\mathbf{k},\omega=0)$ for a charge order with $\bf{2p_{F}}$ and $\bf{q_{S}}$ ordering vectors. We clearly see a gap opening in the AN zone. The magnitude as well as the form factor of the order parameters are the same than in Fig. \ref{allfig}.}
\end{figure}
\end{section}
\newpage
\begin{section}{Electronic structure of a checkerboard CDW order.}
Considering a checkerboard CDW order with an axial ordering vector $q_{cdw}=(q_{x},0)$ and  $q_{cdw}=(0,q_{x})$ with $q_{x}=0.3\pi$, we find that the gap closes from high energies values i.e. from "above" (see Fig. \ref{bandqx} and \ref{gapqx}) as emphasized in \cite{Lee14}.
\begin{figure}[!h]
\includegraphics{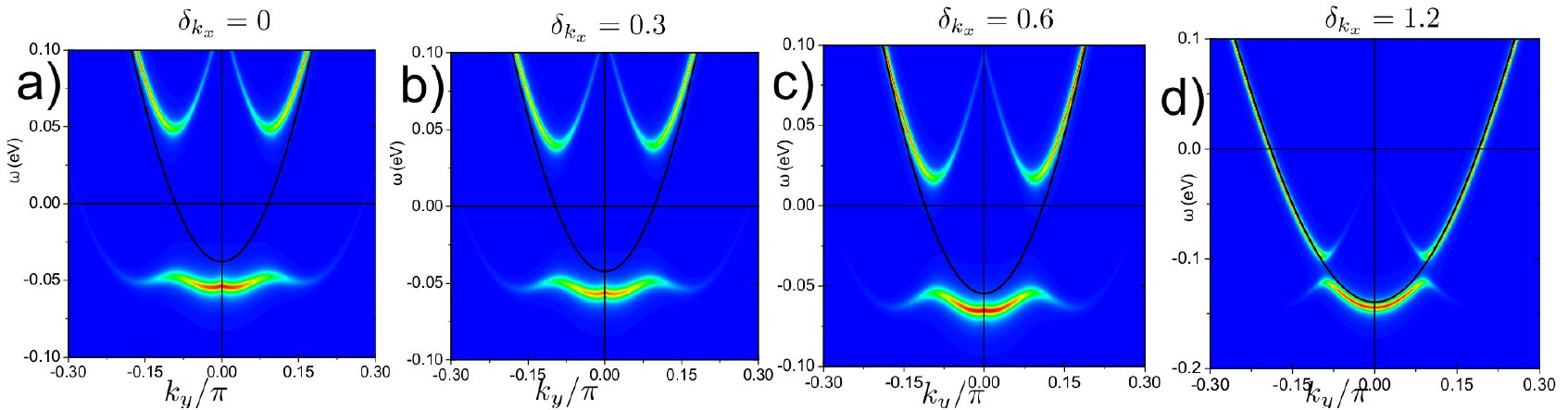}
\caption{\label{bandqx}(Color online) Spectral weight $A(\mathbf{k},\omega)$ as a function of the frequency for a CDW order with a $\bf{q_{cdw}}$ ordering vector at a) $\delta_{k_{x}}=0$, a) $\delta_{k_{x}}=0.3$, a) $\delta_{k_{x}}=0.6$ and a) $\delta_{k_{x}}=1.2$. We clearly see a gap opening in the AN zone. The magnitude as well as the form factor of the order parameters are the same than in Fig. \ref{allfig}.}
\end{figure}
\begin{figure}[!h]
\includegraphics{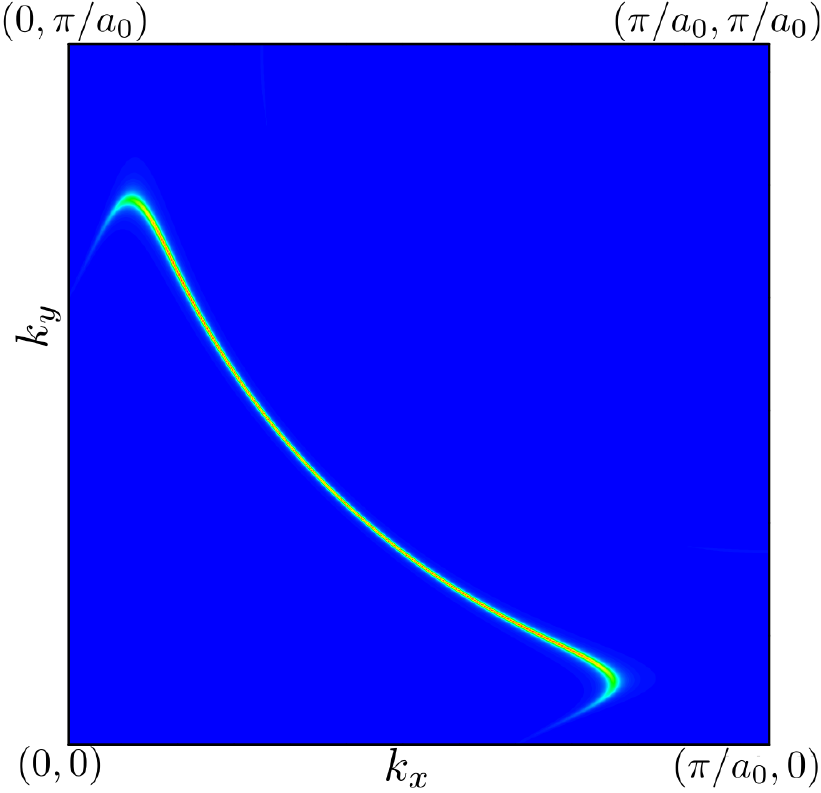}
\caption{\label{gapqx}(Color online) Spectral weight at the Fermi level $A(\mathbf{k},\omega=0)$ for a charge order with a $\bf{q_{cdw}}$ ordering vector. We clearly see a gap opening in the AN zone. The magnitude as well as the form factor of the order parameters are the same than in Fig. \ref{allfig}.}
\end{figure}
\end{section}
\newpage
\begin{section}{Electronic structure of a CDW with a $\bf{2p_{F}}$ ordering vectors}
In this section, we gives some precision about the CDW with a $\bf{2p_{F}}$ ordering vector. The goal of this section is to emphasize the specific role of each ordering vectors contributing to the RES order. Considering a charge order with a $\bf{2p_{F}}$ ordering vectors, a gap opens at the Fermi surface (see Fig. \ref{band2pf}) in the AN zone and the Fermi arcs form (see Fig. \ref{gap2pf}). We clearly see that the gap closes from negative energies i.e. from "below" as observed experimentally \cite{He11,Yoshida:2012kh}.
\begin{figure}[!h]
\includegraphics{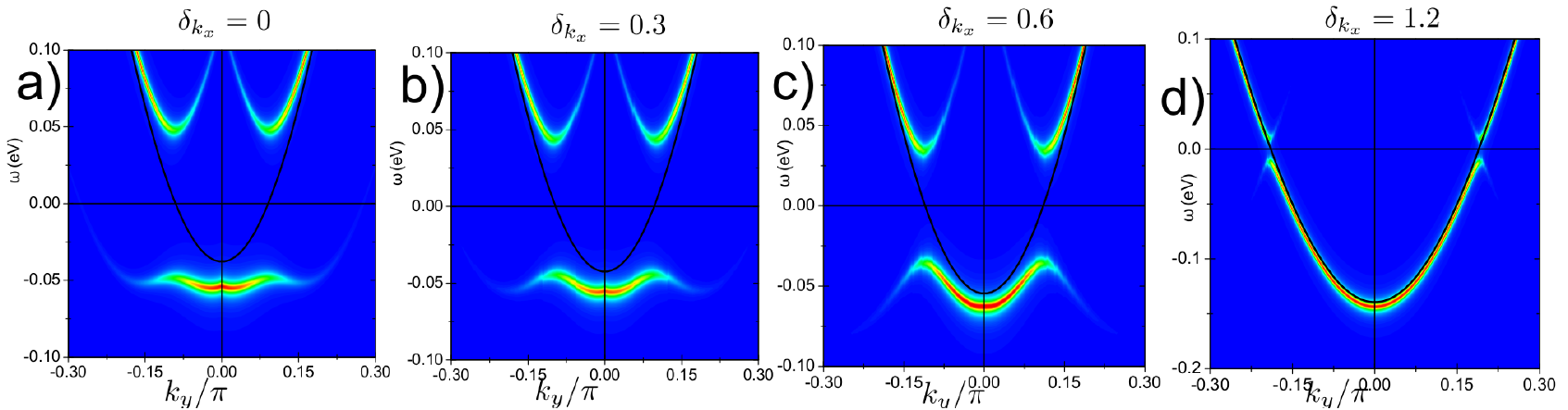}
\caption{\label{band2pf}(Color online) Spectral weight $A(\mathbf{k},\omega)$ as a function of the frequency for a CDW order with a $\bf{2p_{F}}$ ordering vector at a) $\delta_{k_{x}}=0$, a) $\delta_{k_{x}}=0.3$, a) $\delta_{k_{x}}=0.6$ and a) $\delta_{k_{x}}=1.2$. We clearly see a gap opening in the AN zone. The magnitude as well as the form factor of the order parameters are the same than in Fig. \ref{allfig}.}
\end{figure}
\begin{figure}[!h]
\includegraphics{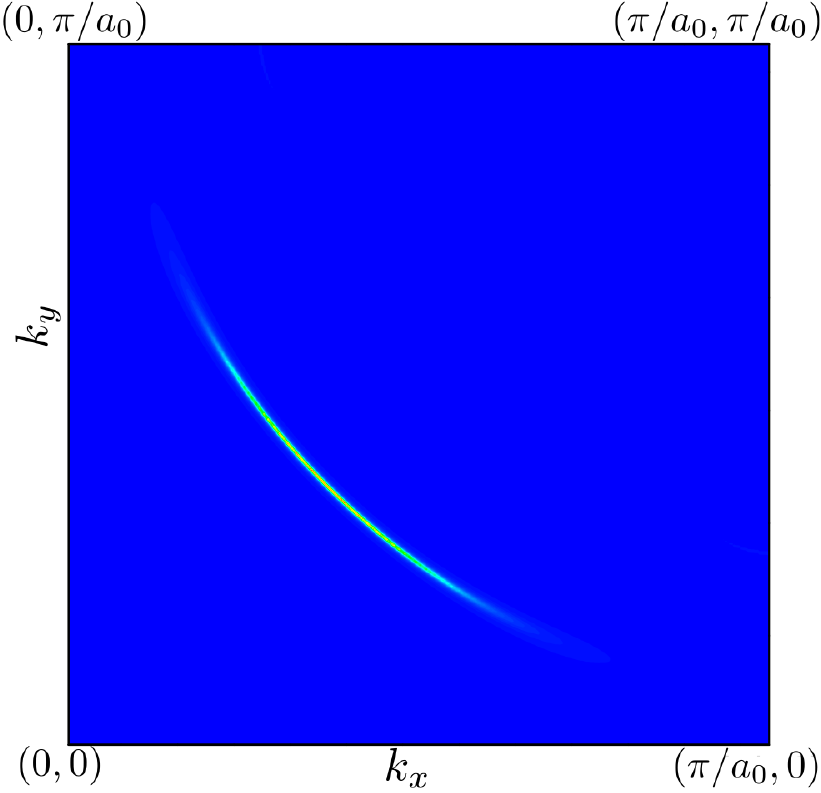}
\caption{\label{gap2pf}(Color online) Spectral weight at the Fermi level $A(\mathbf{k},\omega=0)$ for a charge order with a $\bf{2p_{F}}$ ordering vector. We clearly see a gap opening in the AN zone. The magnitude as well as the form factor of the order parameters are the same than in Fig. \ref{allfig}.}
\end{figure}
\end{section}
\newpage
\begin{section}{Electronic structure of a CDW with a $\bf{q_{S}}$ ordering vector}
In this section, we gives some precision about the CDW with a $\bf{q_{S}}$ ordering vector. A charge order with a $\bf{q_{S}}$ ordering vectors opens a gap in the first BZ but the whole AN zone is not entirely affected (see Fig. \ref{gaphs}). Note that the Fermi surface obtained by the hybridization with $\bf{q_ {S}}$ vector is shifted as regards to the bare Fermi surface allowing the back bending vector $\mathbf{k_{G}}$ to be greater than $\mathbf{k_{F}}$.
\begin{figure}[!h]
\includegraphics{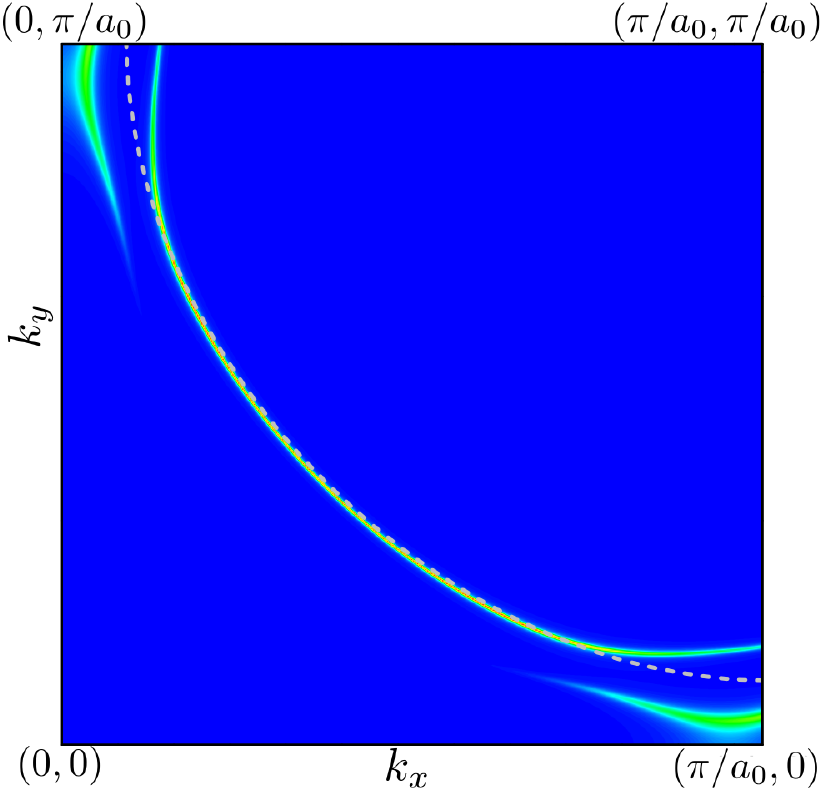}
\caption{\label{gaphs}(Color online) Spectral weight at the Fermi level $A(\mathbf{k},\omega=0)$ for a charge order with a $\bf{q_{S}}$ ordering vector. The bare Fermi surface is presented in dashed line. This ordering vector does not allow the opening of a gap in the whole AN zone. The magnitude as well as the form factor of the order parameters are the same than in Fig. \ref{allfig}.}
\end{figure}
\end{section}
\end{document}